\documentclass[aps,prb,a4paper,twocolumn,showpacs,showkeys,floatfix, superscriptaddress]{revtex4-2}

\usepackage{graphicx}
\usepackage{color}
\usepackage{amsmath}
\usepackage{multirow}

\newcommand{\vect}[1]{\mathbf{#1}}
\newcommand{\RbFe}{RbFe(MoO$_4$)$_2$}
\newcommand{\RbKFe}[2]{Rb$_{#1}$K$_{#2}$Fe(MoO$_4$)$_2$}
\newcommand{\abs}[1]{\left|#1\right|}
\begin{document}

\title{Stabilization of the collinear plateau phase by thermal fluctuations in the disordered triangular lattice antiferromagnet Rb$_{(1-x)}$K$_{x}$Fe(MoO$_4$)$_2$}

\author{V. N. Glazkov}
\email{glazkov@kapitza.ras.ru}
\affiliation{ P.L. Kapitza Institute for Physical Problems, RAS, Kosygin str. 2, Moscow 119334, Russia}

\author{I. A. Krastilevskiy}
\affiliation{ P.L. Kapitza Institute for Physical Problems, RAS, Kosygin str. 2, Moscow 119334, Russia}

\begin{abstract}
The triangular lattice antiferromagnet \RbFe{} orders antiferromagnetically in a planar 120$^\circ$-structure below $T_\textrm{N}\approx 4$~K. A striking feature of \RbFe{} magnetic phase diagram is the presence of collinear ``1/3-plateau'' magnetic phase, which is stabilized by thermal and quantum fluctuations at magnetization $M\approx \frac{1}{3} M_\textrm{sat}$. Static disorder caused by impurities is predicted to  act against the effect of fluctuations and to suppress collinear plateau phase (Maryasin and  Zhitomirsky, PRL \textbf{111}, 247201 (2013)). Balance between  ``dynamic'' thermal and quantum fluctuations and ``static'' impurity-induced disorder is temperature-sensitive, which allows thermal fluctuations to take over the effect of static disorder and leads to the revival of the fluctuation-stabilized ``1/3-plateau'' phase on heating.  Here we present  experimental results directly  confirming this prediction and demonstrating  re-establishment of the plateau-like phase in the diluted  \RbKFe{(1-x)}{x} sample at moderate dilution level $x=15$\% on heating as the effect of thermal fluctuations increases.
\end{abstract}

\date{\today}



\maketitle
\section{Introduction}

The triangular lattice Heisenberg antiferromagnet is one of the cornerstone problems of frustrated magnetism \cite{anderson, frazecas, miyashita1, miyashita2, collins}. Competition of the exchange couplings along the bonds of a triangular lattice prevents a ``trivial'' up-down-up-down N\'{e}el ordering pattern and the choice of the ground state is governed by order-by-disorder effects caused by quantum and thermal fluctuations. As a result, zero-field magnetic structure of a semi-classical two-dimensional triangular lattice Heisenberg antiferromagnet is a chiral three-sublattice ``120${}^\circ$-structure'' with the net spin per triangle being zero and three spin vectors on the same triangle being aligned at 120${}^\circ$ angle to each other. The chirality of the 120${}^\circ$-structure leads to the multiferroic properties of triangular lattice antiferromagnets (see, e.g. \cite{korean,kenzelman} ) and importance of fluctuations is also highlighted by their impact on spin-wave spectra \cite{zhit-sw}.

Application of magnetic field leads to the non-trivial magnetic phase diagram of the triangular lattice  Heisenberg antiferromagnet \cite{ kawa,chub,kawa2,mila}: due to the thermal and quantum fluctuations favoring the most collinear phase through order-by-disorder mechanism a collinear up-up-down (or ``uud'') phase with the net magnetization $\frac{1}{3}M_\textrm{sat}$ is stabilized over a certain range of magnetic fields. This feature is observed on magnetization curves $M(B)$ as a plateau of magnetization at 1/3 of saturation level (and, thus, as frequently referred to as a ``1/3-plateau'' phase). Being a characteristic feature of a triangular lattice antiferromagnet, the magnetization plateau was observed in different systems (see, e.g., \cite{smirnov2007,plateau1,plateau2}).

Stability of this fluctuation-stabilized phase is a subject of separate interest. Introduction of disorder, e.g.,  site-disorder through the decimation of the magnetic lattice or a bond-disorder through the random modification of the  exchange bonds, is one of the ways to act upon the spin system. Theoretical analysis and Monte-Carlo simulations of Ref.~\cite{maryasin} revealed that the effect of impurities producing  a static (or ``frozen'') disorder, is opposite to the effect of the ``dynamic'' disorder created by fluctuations: while fluctuations favor the most collinear phase, thus stabilizing the ``uud'' state, ``frozen'' disorder prefers the most non-collinear phase suppressing the plateau phase of the triangular antiferromagnet. The most interesting prediction of Ref.~\cite{maryasin} is that a balance between static disorder and fluctuations is temperature-dependent: at certain level of disorder (e.g., decimation of 5\% of the lattice sites in the model of Ref.~\cite{maryasin}) ``frozen'' disorder overcomes the low-temperature effect of quantum fluctuations, but thermal fluctuations can, in turn, become dominant as the temperature increases. Thus, a suppression of the collinear plateau phase at low temperature and revival of this phase on heating are predicted by theory. Here, we report experimental results on the magnetic phase diagrams  of diluted \RbKFe{1-x}{x}{}, demonstrating validity of this prediction.

\RbFe{} is a well established experimental realization of a triangular lattice antiferromagnet. Its structure features Fe$^{3+}$ ($S=5/2$) triangular layers orthogonal to the $C_3$ symmetry axis separated by Rb ions and MoO$_4$ complexes (intra-layer Fe-Fe distance is 5.67~\AA, while inter-layer Fe-Fe distance is 7.49~\AA) \cite{klimin,inami}. The magnetic properties of this quasi-2D compound were studied using  magnetization and specific heat measurements \cite{svistov2003, smirnov2007, smirnov-ufn}, ESR and NMR techniques \cite{svistov2003, smirnov2007,svistov-NMR-jetp,sakhratov-jetp,smirnov-jpcs}, and neutron scattering \cite{inami, kenzelman,white}.

Spin subsystem of \RbFe{} can be described within each two-dimensional layer by the following Hamiltonian \cite{white}:

\begin{equation}\label{eqn:ham}
{\cal H}=J \sum_{\langle i,j\rangle}  {\hat{\vect{S}}}_i{\hat{\vect{S}}}_j+D \sum_i {\hat{S}}_{i,z}^2+g\mu_\textrm{B} \vect {B}\sum_i{\hat{\vect{S}}}_i
\end{equation}

\noindent here the intra-layer coupling $J=0.086$~meV and planar (XY) anisotropy constant $D=0.027$~meV. The inter-layer coupling constant is  $J'=0.0007$~meV.

Pure \RbFe{} orders antiferromagnetically at $T_\textrm{N} \approx 3.8$~K. At zero field the ordered atomic magnetic moments are confined to the plane $\perp C_3$ and form a three-sublattice  ``120$^\circ$-structure'' within each layer. Minute distortions of inter-layer couplings caused by high-temperature structural transitions \cite{klimin} lead to the incommensurate rotations of these 120$^\circ$-ordered spin structures along the $C_3$ axis \cite{kenzelman,white,svistov-NMR-jetp}. Magnetic phase diagrams of pure \RbFe{} were studied both for $\vect{B}\perp C_3$ \cite{smirnov2007,kenzelman} and for $\vect{B}||C_3$ \cite{sakhratov-jetp}.

In the case of present interest, $\vect{B}\perp C_3$, applying a magnetic field leads to a staircase of phase transitions: first, an incommensurate-commensurate phase transition locks inter-layer ordering pattern \cite{white,kenzelman}, second, a collinear ``uud'' phase is stabilized around 1/3 of the saturation field, then a high-field commensurate-incommensurate transition is observed, and finally spin system saturates at $H_\textrm{sat} \simeq 17$~T.

``Static'' structural disorder can be introduced to this compound by non-magnetic dilution: it is possible to grow single crystals of \RbKFe{1-x}{x}{} with $x\leq15$\% \cite{smirnov2017}.  Introduction of the K ions instead of Rb creates bond disorder instead of site disorder considered in Ref.~\cite{maryasin}, however, similar effect is expected.  The effect of plateau phase revival was sought experimentally by studying the potassium-diluted \RbKFe{1-x}{x}{} for $\vect{B}\perp C_3$  by magnetization measurements, ESR and NMR techniques  \cite{smirnov-jpcs, smirnov2017, sakhratov}. The experimental results were not entirely decisive: the magnetization study of Ref.~\cite{smirnov2017} demonstrated that a characteristic decrease of the $dM/dB$ derivative disappears at low temperature for $x=15$\% \RbKFe{1-x}{x}{} but did not study phase boundaries in details; the ESR study of Refs.~\cite{smirnov-jpcs,smirnov2017} proved that the  low temperature ordered phase for $x=15$\% sample is different from that of pure \RbFe{} but did not follow temperature evolution of this phase; the NMR study of Ref.~\cite{sakhratov} concluded that a ``fan'' structure (see next section for phases description) with some inter-layer disorder is formed in $x=15$\% \RbKFe{1-x}{x}{} sample, but this study  was carried out along a single line on the $(BT)$ plane, which, as we will discuss below, can miss the plateau-like phase.

Below we report the results of detailed study of magnetic phase diagrams of \RbKFe{1-x}{x}{} with different dilution levels. These results demonstrate that a plateau (or plateau-like) phase is formed at a dilution level $x=15$\% in a well-defined area of a $(BT)$ plane providing an evidence for the revival of the plateau phase under the increasing influence of thermal fluctuations.

\section{Experimental details and samples}
Samples of pure \RbFe{} and K-diluted \RbKFe{(1-x)}{x}{} with $x=7.5, 15$\% were kindly provided by Dr. A.Ya.Shapiro and Prof. L.N.Demianets (Institute of crystallography RAS). Same samples or samples from the same growth batches were used in Refs.~\cite{smirnov-jpcs,smirnov2017,sakhratov,svistov-NMR-jetp,sakhratov-jetp}.

The magnetization study of Ref.~\cite{smirnov2017} confirms antiferromagnetic ordering in all \RbKFe{(1-x)}{x}{} samples with $x\leq 15$\%, with the low-field  N\'{e}el temperature monotonically decreasing from 3.8~K to 3.0~K as the  potassium content increases. High-field (up to 20 T) magnetization curves demonstrate well-defined drop in the differential susceptibility $dM/dB$ at the ``1/3-plateau'' phase for  pure \RbFe{}. This decrease in the slope of the magnetization curve became smeared in the K-diluted samples and almost disappeared at the highest dilution level of 15\%.

Low-temperature antiferromagnetic resonance study \cite{smirnov2017,smirnov-jpcs} demonstrated that electron-spin-resonance absorption patterns are completely different in pure \RbFe{} and the 15\% K-diluted sample, unambiguously proving that the low temperature (1.3~K) magnetic ordered states are different in these samples.

In the present paper we performed a detailed study of the magnetic phase diagrams of \RbKFe{(1-x)}{x}{} samples with $x=0,~7.5,~15$\% using magnetic torque measurements technique with the home-made capacitance torque magnetometer.

The experimental cell was a flat capacitor (plate area about $4 \times 4$~mm$^2$, capacitance $C_0 \simeq 1$~pF) with one plate made of 50~$\mu$m flexible bronze cantilever and the second stationary. The sample was glued to the flexible plate, and the magnetic field was applied parallel to the cantilever. Transverse magnetization of the sample $\vect{M}_\perp$ produced  the magnetic torque $\vect{T}=\vect{M}_\perp \times \vect B$, which was compensated by the elastic forces of the bent cantilever. The cantilever bending resulted in a change of capacitor inter-plate distance $d$, which yielded measurable change of the capacitance.

Assuming that the capacitor plates remain parallel, and that the elastic force is proportional to the change of the inter-plates distance $\delta d$, the net transverse magnetic moment of the sample can be expressed as
\begin{equation}\label{eqn:torque}
M_\perp B\propto \delta d \propto \frac{\abs{C-C_0}}{C C_0}
\end{equation}

\noindent where $C(B,T)$ is a measured capacitance, and $C_0$ is the initial zero-field capacitance.

The experimental cell was mounted on He-4 or He-3 vapor pumping cryostats equipped with superconducting cryomagnets. The base temperature  of the He-3 cryostat was 0.38~K. The maximal magnetic field was 7~T for the He-4 cryostat and 12~T for the He-3 cryostat. Capacitance $C(B,T)$ was measured with a capacitance bridge as a function of magnetic field at fixed temperature (B-scans), or as a function of temperature at fixed field (T-scans).

Transverse magnetization of the non-ferromagnetic sample arises as a result of susceptibility anisotropy when the field direction deviates from the crystal symmetry axes. This effect is enhanced near the magnetic reorientation transitions where  the susceptibility tensor undergoes strong changes. Additional effects leading to the change of the cell capacitance are the sample shape effect resulting in the deviation of the  sample magnetization vector $\vect {M}$ from the direction of the applied field  $\vect{B}$ and  the magnetic forces due to the field gradients at the sample position.

Thus, the total torque-meter response is an unknown function of transverse and longitudinal magnetization of the sample. However, as both of them change at a magnetic phase transition, magnetic torque measurement has proved to be a very sensitive technique to detect magnetic phase transitions and to study  magnetic phase diagrams \cite{glazkov-pyr1,glazkov-pyr2} or other physical effects related to the magnetization change, e.g. de Haas-van Alphen effect \cite{sheikh}.

The determination of magnetic phase transition in magnetic torque measurements relies upon qualitative changes in torque-meter response, which usually take form of the sharp kinks.
We will use response function $R(B,T)$ proportional to the ``ideal'' transverse magnetization~\eqref{eqn:torque} as a representation of the measured data

\begin{equation}
\label{eqn:response}
R(B,T)=\frac{1}{B} \abs{\frac{1}{C(B,T)}-\frac{1}{C_0}}
\end{equation}
\noindent  Data below approximately 1.5~T were excluded from $R(B,T)$ calculations, since the small capacitance change is strongly affected by measurement noise. When divided by small field $B$, this yields senselessly noisy data. To make  $R(B,T)$ curves at higher fields  more suitable for the presentation and for the derivatives $dR/dB$, $dR/dT$ analysis we smoothed them with the spline.

\section{Expected torque-meter response: Three-sublattice model and magnetic phases. }
\begin{figure}[th]
\centering
  \includegraphics[width=\columnwidth]{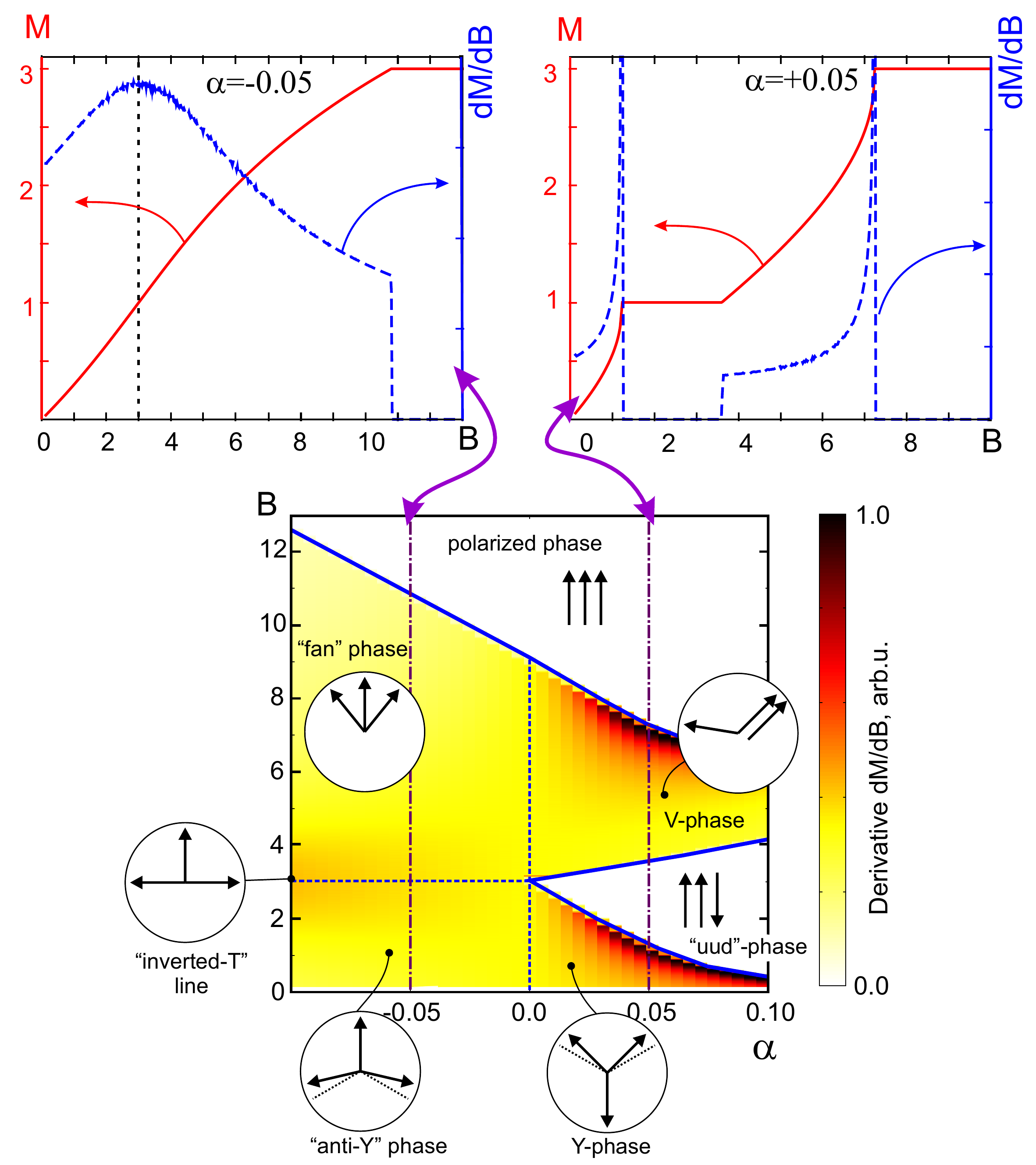}\\
  \caption{(color online) Main panel: Phase diagram of the three-sublattice toy model (see Eqn.~\eqref{eqn:energy}) for the in-plane magnetic field as a function of the biquadratic coupling parameter $\alpha$. Color map on the main panel shows derivative $dM/dB$. Arrows schematically show mutual orientation of sublattices in different magnetic phases. Thick solid lines mark transitions to collinear plateau phase (``uud''-phase) and to the polarized high-field phase. Dashed lines mark cross-over from ``anti-Y'' to ``fan'' phase at $M=\frac{1}{3} M_\textrm{sat}$ for $\alpha<0$ and  ``line of degeneracy'' at $\alpha=0$. Dash-dotted lines at $\alpha=\pm0.05$ show positions of the magnetization field scans shown on upper panels. Upper panels: model $M(B)$ curves and their field derivatives $dM/dB$ computed within the  three-sublattice toy model described by Eqn.~\eqref{eqn:energy}  for  $\alpha=-0.05$ (left) and $\alpha=+0.05$ (right).}\label{fig:model}
\end{figure}

\begin{figure}[th]
\centering
  \includegraphics[width=\columnwidth]{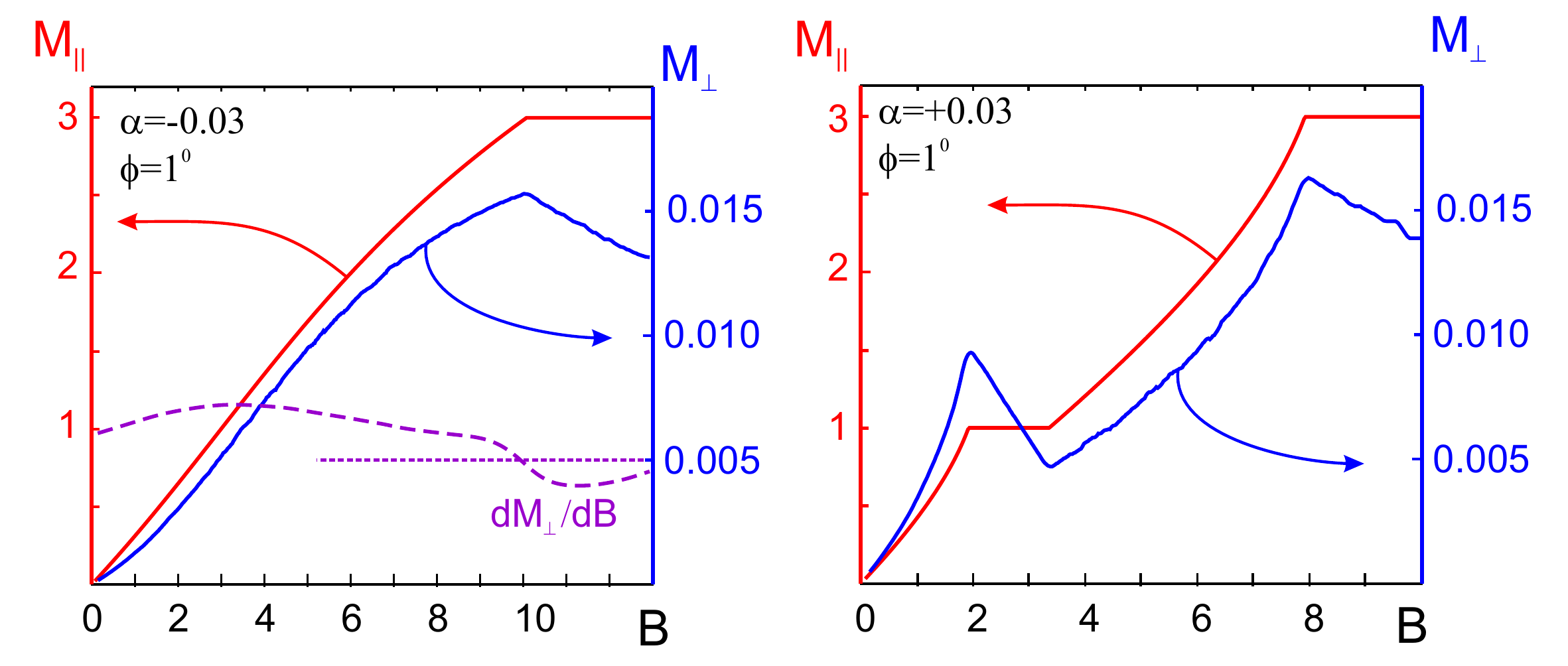}\\
  \caption{(color online) Field dependence of the longitudinal and transverse magnetization for the three-sublattice toy model described by Eqn.~\eqref{eqn:energy} in slightly out-of-plane magnetic field (1$^\circ$ out of the easy plane). Left panel: $J=1$, $D=0.3$, $\alpha=-0.03$. Right panel: $J=1$, $D=0.3$, $\alpha=0.03$. Dashed line on the left panel shows derivative $dM_\perp/dB$, horizontal dashed line indicates the derivative's zero value.}\label{fig:torquemodel}
\end{figure}

To understand qualitatively what kind of features in torque-meter response $R(B,T)$ should be associated with the ``1/3-plateau'' phase of a triangular lattice antiferromagnet we performed simple modeling of $M_\perp (B)$ curves. We considered single 2D layer and assumed three-sublattice ordering within it.  We assumed sublattices' total spins $\vect{S}_{1,2,3}$ to be unit vectors and considered energy of three-sublattice antiferromagnet in a form

\begin{eqnarray}
E&=&\frac{J}{2}\left(\vect{S}_1\vect{S}_2+\vect{S}_2\vect{S}_3+\vect{S}_1\vect{S}_3 \right)+D\left(S_{1z}^2+S_{2z}^2+S_{3z}^2\right)-\nonumber\\
&&-\alpha\left[\left(\vect{S}_1\vect{S}_2\right)^2+\left(\vect{S}_2\vect{S}_3\right)^2+\left(\vect{S}_1\vect{S}_3\right)^2\right]-\nonumber\\
&&-\frac{1}{6}\vect{B}\left(\vect{S}_1+\vect{S}_2+\vect{S}_3\right)\label{eqn:energy}
\end{eqnarray}

\noindent here $J$ is an exchange coupling constant ($J=1$ for the modeling purpose), $D>0$ is an easy plane anisotropy constant (we use $D/J=0.3$ which is closed to the known ratio for \RbFe{} \cite{white}). The positive biquadratic coupling parameter $\alpha>0$ describes effect of spin fluctuations towards stabilization of the collinear spin structure  \cite{maryasin,biquad,smirnov2017} (we take $\alpha/J=0.03$ for torque-meter response modeling to produce  reasonably broad plateau region), negative biquadratic coupling parameter $\alpha<0$ favors non-collinear spin structures and phenomenologically mimics the effect of ``frozen'' disorder \cite{maryasin}.  Factors $1/2$ and $1/6$ at exchange and Zeeman terms are introduced for compatibility of the sublattices model with the  lattice sum \eqref{eqn:ham} as each bond is shared by two triangles and each site is shared by six triangles on the triangular lattice. This oversimplified toy model does not pretend on physically correct description of temperature effects (which requires a more accurate treatment, see, e.g., \cite{mila}), however it gives useful insight on the expected change of the transverse magnetization in our experiment and correctly reproduces magnetic phases.

Minimization of this energy with respect to the  $\vect{S}_{1,2,3}$ vectors orientation allows to determine the net magnetization vector $\vect{M}$ (Fig.~\ref{fig:model}). For the field applied within the easy plane three-sublattice model \eqref{eqn:energy} saturates at $B_\textrm{sat}=9 J$ at $\alpha=0$. The $\alpha=0$ case is highly degenerate: all sublattices' configurations with the same net magnetic moment have the same energy. Non-zero biquadratic term lifts this degeneracy.

At $\alpha>0$ low-field phase is a symmetric Y-phase with upper arms of  ``Y'' bending towards field direction. At certain field arms of ``Y'' became parallel and a collinear ``1/3-plateau'' phase is stabilized in the vicinity of $B_\textrm{sat}/3$. The noncollinear  V-phase with two parallel sublattices appears on further increase of the field, which finally folds into spin-polarized phase at saturation field.

At $\alpha<0$ low-field phase is a symmetric ``anti-Y'' phase with lower arms of ``anti-Y'' spreading towards the field, this sublattices configuration continuously transforms to the symmetric ``fan'' phase with all sublattices being non-collinear. There is no magnetization plateau for $\alpha<0$, however, there is a maximum of $dM/dB$ at $B=3J$, which corresponds to the ``inverted-T'' line on the phase diagram of this simple model: along this line $M=M_\textrm{sat}/3$, one sublattice is aligned along the field and two other sublattices are orthogonal to the field.

For the in-plane magnetic field no transverse magnetization appears in this model.  However, if the magnetic field is directed slightly out of the plane (we took 1$^\circ$ out-of-plane deviation for the modeling) transverse magnetization $M_\perp$ appears, its field dependence closely mimics the $M_\parallel(B)$ behavior (Fig.~\ref{fig:torquemodel}). For the fluctuation-governed regime $\alpha>0$  we conclude that the expected features of the torque-meter response are: (i) interval of a negative (or at least reduced) slope on $M_\perp(B)$ curve over the plateau region and (ii) kink at the saturation field. For the ``frozen''-disorder governed regime $\alpha<0$ only kink at the saturation field is expected and probably cross-over-like maximum of $dM_\perp/dB$ along the ``inverted-T'' line.

\section{Experimental results}
\subsection{Magnetic phase diagram of pure \RbFe{}: proof of method reliability}
\begin{figure}[th]
\centering
  \includegraphics[width=\columnwidth]{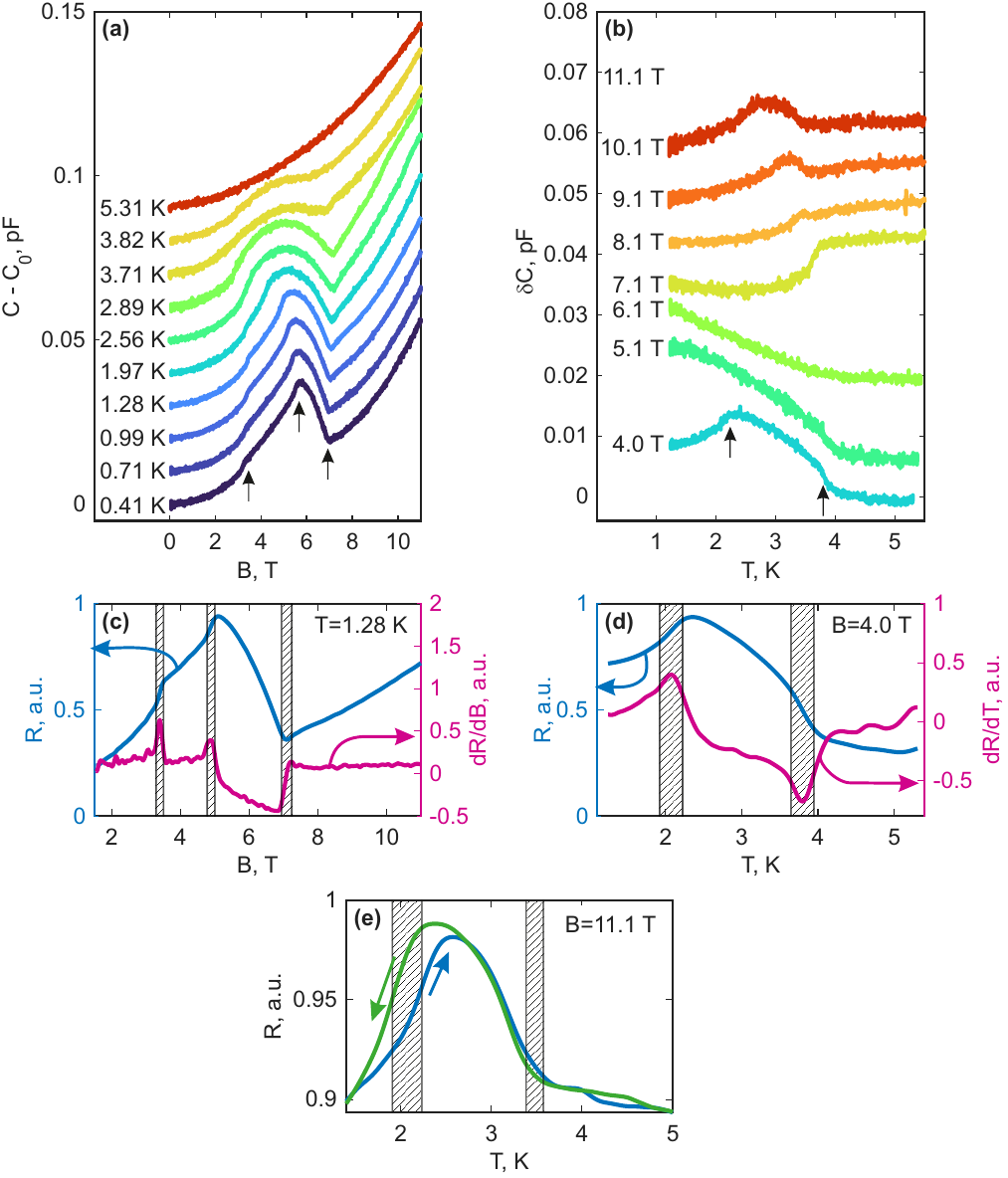}\\
  \caption{(color online) Determination of the phase transitions for the pure \RbFe{}.  ((a) and (b)) Examples of the raw data (cell capacitance change) during B-scans (a) and T-scans (b) at different temperatures and fields, arrows mark features interpreted as a phase transition. ((c) and (d)) Representative examples of the  response function $R(B,T)$ (see Eqn.~\eqref{eqn:response}) and its derivatives $dR/dB$ and $dR/dT$, $R(B,T)$ curves are smoothed with spline for derivative calculation (see text). (e) Example of the response function $R(B,T)$ showing hysteresis behavior at  the first-order phase transition between  incommensurate and commensurate phases, direction of the temperature sweep is shown by arrows.  Vertical hatched stripes in panels (c), (d) and (e) show estimated error margins for the phase transition field or temperature.}\label{fig:pure-data}
\end{figure}

\begin{figure}[th]
\centering
  \includegraphics[width=\columnwidth]{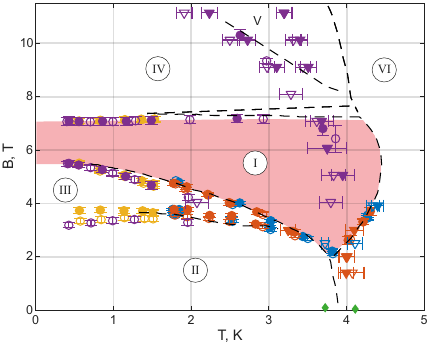}\\
  \caption{(color online) Magnetic phase diagram of pure \RbFe{}. Symbols mark phase transitions determined from the present torque measurements (open circles --- B-scans on decreasing field, closed circles --- B-scans on increasing field, open triangles --- T-scans on cooling, closed triangles --- T-scans on heating). Zero-field transition temperatures (green diamonds) are taken from Ref.~\cite{smirnov2007}, two values of zero-field $T_\textrm{N}$ correspond to different sets of samples (see text). Dashed lines --- known phase boundaries from Ref.~\cite{smirnov2007}. Phases are marked as follows: (I) ``1/3-plateau'' phase, (II) incommensurate low-field phase, (III) commensurate low-field phase, (IV) commensurate high-field phase, (V) incommensurate high-field phase, (VI) paramagnetic phase. Shaded area  highlights the ``1/3-plateau'' phase.}\label{fig:pure}
\end{figure}

The phase diagram of pure \RbFe{} is well known \cite{smirnov2007,kenzelman}: low temperature saturation field reaches 18~T at 1~K, ``1/3-plateau'' (or ``uud'') phase is observed at a certain field range below 7~T, additionally two commensurate-incommensurate transitions are observed in the phases below and above the magnetization plateau, these transitions are related to the inter-layer ordering patterns \cite{kenzelman}. We studied pure \RbFe{} samples with torque-meter as a proof of concept to demonstrate that the plateau phase boundaries can be reliably determined by this technique.

Torque-meter response curves (see Fig.~\ref{fig:pure-data}) reveal well defined kinks both on field- and temperature-scans (B-scans and T-scans). Noticeably, low-temperature field scans demonstrate negative slope on $R(B)$ curve, as expected for the plateau phase (compare with Fig.~\ref{fig:torquemodel}). Besides  the negative slope fragment $R(B)$ curve also features a low-field kink at the commensurate-incommensurate transition. T-scans (see Fig.~\ref{fig:pure-data}) are also informative: there is a clear kink at the paramagnetic phase boundary, at the transition from the plateau phase  and at the high-field commensurate-incommensurate transition. First-order phase transitions from incommensurate to commensurate phase are accompanied by clearly visible hysteresis both on B-scans and on T-scans, while second-order transitions to and from the plateau phase and to the paramagnetic phase are hysteresis-free within the experimental accuracy.

Pinning down the phase transition field and temperature with torque-meter technique is, to some extent, a matter of convention (which, of course, has to be accounted for by the appropriate error margins). Magnetization modeling (Fig.~\ref{fig:model}) predicts a sharp peak in $dM/dB$ on approaching the plateau phase from below. We found that the derivatives $dR/dB$ and $dR/dT$ of the torque-meter response function $R(B,T)$ do provide sharp anomalies. In the case of a well-defined maximum or minimum on the  $dR/dB$ or $dR/dT$ curve we took the position of this extremum  as the transition point. In the case of kink-like features (where the derivative promptly changes between two values without a well-defined extremum) we took the visual position of the kink as the transition point. Error margins were determined as the derivative peak width or the width of the prompt derivative change interval.

Resulting phase diagram is shown in Fig.~\ref{fig:pure}. Positions of the identified features agree well with the known phase boundaries \cite{smirnov2007}. Hysteresis behavior is observed along the first-order transition lines. Certain  discrepancy on the high-field data paramagnetic phase boundary is a  known sample quality issue (the effect of trace Al impurities due to the growth procedure shifts  the samples'  N\'{e}el temperature but practically does not affect the magnetic phases, see Ref.~\cite{smirnov2017}).

Summing up, our measurements demonstrate that the well-defined features of the  torque-meter response function $R(B,T)$ really mark the phase boundaries in \RbFe{}. Our data on pure \RbFe{} also provides additional data on the  phase diagram of this compound below 1 K.

\subsection{Magnetic phase diagram of \RbKFe{(1-x)}{x}{} at low dilution level $x=7.5$\%: robustness of the plateau phase at low dilution level}
\begin{figure}[th]
\centering
  \includegraphics[width=\columnwidth]{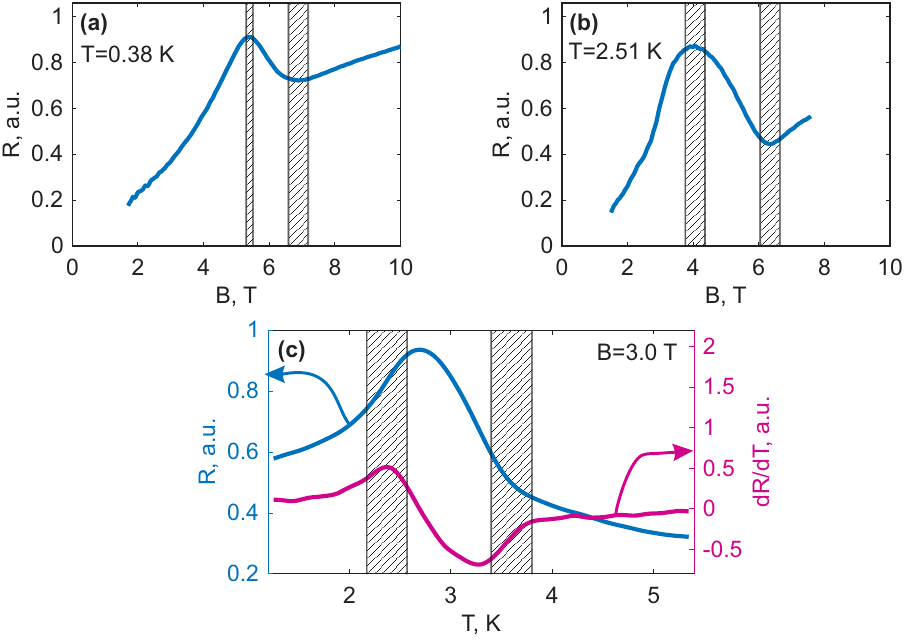}\\
  \caption{(color online) Examples of the torque-meter response function $R(B,T)$ for \RbKFe{(1-x)}{x}{} sample with $x=7.5$\%. ((a) and (b))  B-scans at different temperatures, (c) T-scan. Vertical hatched stripes indicate error margins for the identified phase transitions.   }\label{fig:7p5data}
\end{figure}

\begin{figure}[th]
\centering
  \includegraphics[width=\columnwidth]{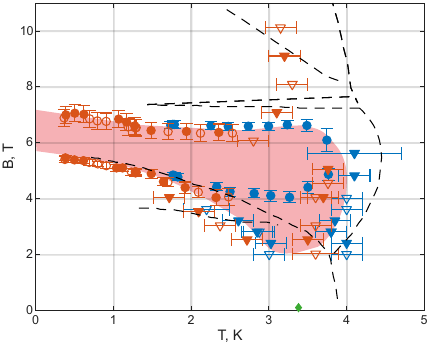}\\
  \caption{(color online) Magnetic phase diagram of \RbKFe{(1-x)}{x}{} at low dilution level $x=7.5$\%. Circles: B-scans (open symbols corresponds to the measurements with decreasing field, closed symbols --- to the measurements with increasing field). Triangles: T-scans (open symbols correspond to the data taken on cooling, closed symbols --- on heating). Green diamond: zero-field N\'{e}el point from magnetization measurements of Ref.~\cite{smirnov2017}. Dashed lines:  phase boundaries for the pure \RbFe{} from Ref.~\cite{smirnov2007}. Shaded area highlights the plateau-like phase.}\label{fig:7p5}
\end{figure}

Examples of the torque-meter response curves for the $x=7.5$\% \RbKFe{(1-x)}{x}{}  sample are shown in Fig.~\ref{fig:7p5data}. Partial substitution of Rb ions by K ions results in smearing of the phase transition features. No clear  features related to the commensurate-incommensurate phase transitions were detected. However, B-scans feature fragments with a negative slope, which is the fingerprint of the plateau phase. Derivative ${dR}/{dB}$ analysis proved indecisive for the smeared transitions, so we took maximum of the $R(B)$ curve as (quite likely, overestimated) lower boundary of the plateau phase and high-field kink as  upper boundary of the plateau phase. For the T-scans, the kink on the high-temperature shoulder of the $R(T)$ curve was used as upper boundary of the plateau phase and derivative peak on the left shoulder was taken as a lower boundary of the plateau phase. Uncertainty in the identification of the smeared phase transition was taken into account as the appropriate error bars.

All results are collected in Fig.~\ref{fig:7p5}. One can see that the phase diagram of the weakly ($x=7.5$\%) diluted \RbKFe{1-x}{x}{} includes an extended area with negative $R(B)$ slope which we identify as a ``1/3-plateau'' phase. This phase is slightly shifted to lower temperatures and to  lower fields compared to the pure \RbFe{}. Lower boundary of the plateau phase at 2.5--3.5~K temperature range was more reliably determined from T-scans, while B-scans (as  marked above) do yield overestimated transition fields. Such a discrepancy is in qualitative agreement with the increasing slope ${dB}/{dT}$ of the low-field  phase transition lines at 2.5--3.5~K temperature range. As a result, T-scans ``cut'' the  phase transition line almost normally to the phase transition line yielding quite sharp features on the response curves, while B-scans approach the phase transition line tangentially yielding very smooth features on the response curves. At low temperatures B-scans ``cut'' the phase transition lines almost normally and all data come into agreement.

Thus, we have measured in detail the  magnetic phase diagram  of \RbKFe{(1-x)}{x}{} at low dilution level $x=7.5$\%. Our results indicate that at low dilution level the plateau phase remains robust to the effect of the ``frozen'' disorder and that quantum and thermal fluctuations are still strong enough to stabilize the collinear ``1/3-plateau'' phase.

\subsection{Magnetic phase diagram of \RbKFe{(1-x)}{x}{} at high dilution level $x=15$\%: suppression   of the plateau phase and its revival under the effect of thermal fluctuations}
\begin{figure}[th]
\centering
  \includegraphics[width=\columnwidth]{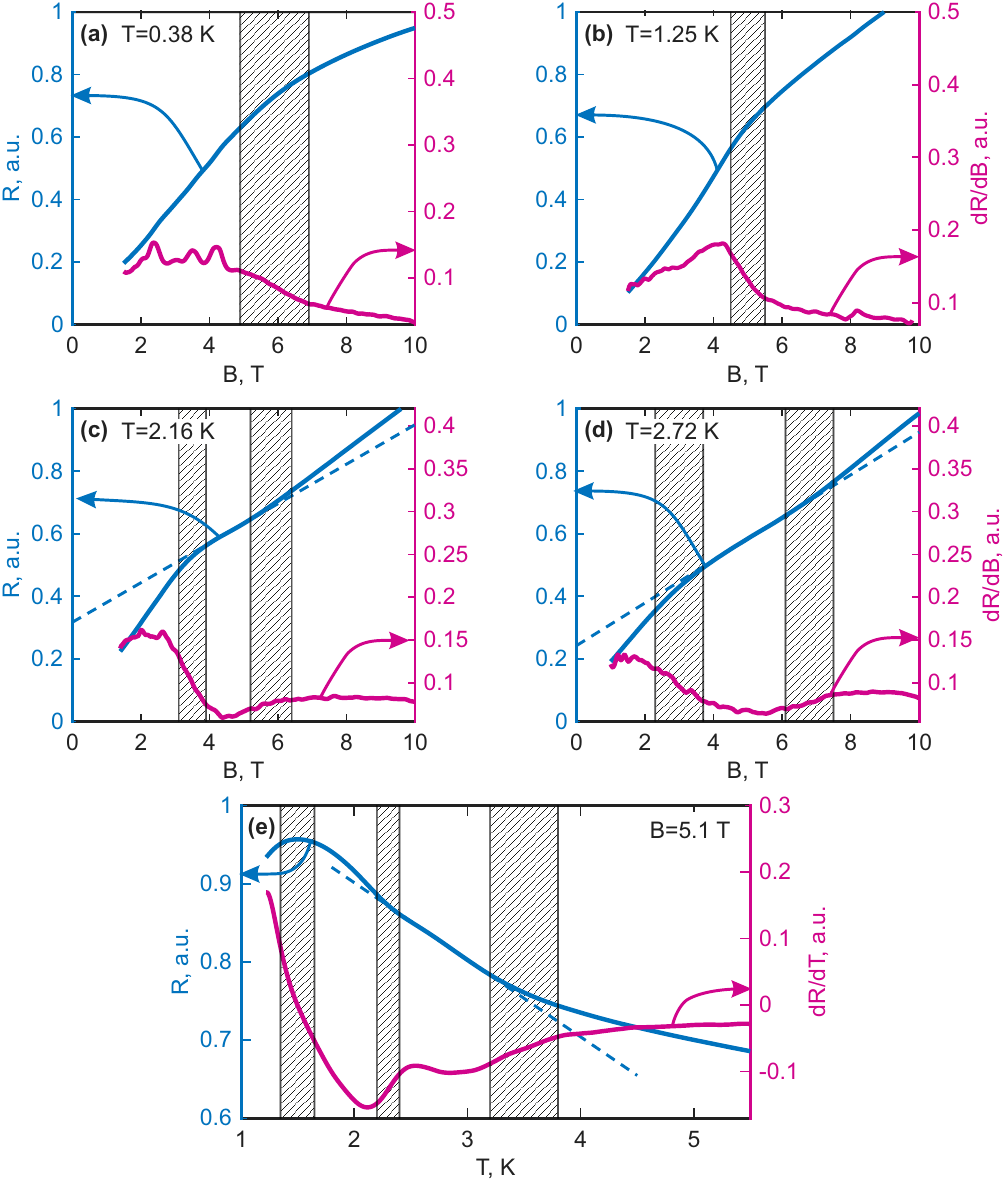}\\
  \caption{(color online) Examples of torque-meter response function $R(B,T)$ for \RbKFe{(1-x)}{x}{} sample with $x=15$\%.
  ((a) and (b)) Low temperature B-scans with \emph{single} kink. ((c) and (d)) Higher temperature B-scans with \emph{two} kinks. (e) T-scan at the selected field of 5.1~T with two kinks.  Dashed straight lines are guides-to-the-eye highlighting the fragments of $R(B,T)$ response curves with changed slope.  Vertical hatched stripes in all panels indicate error margins of the identified phase transitions.  }\label{fig:15-data}
\end{figure}

\begin{figure}[th]
\centering
  \includegraphics[width=0.8\columnwidth]{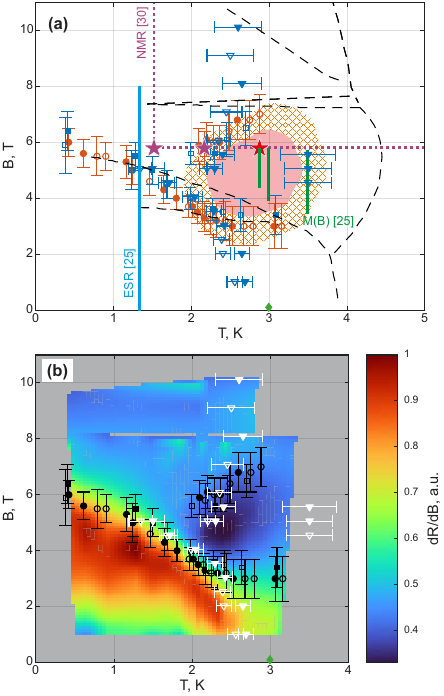}\\
  \caption{(color online) Magnetic phase diagram of \RbKFe{1-x}{x}{} with $x=15$\%. Panel (a) shows identified phase transitions, shaded area highlights limits of the plateau-like phase (hatched border approximately corresponds to the error margins of the identified phase transitions),  dashed lines mark  phase boundaries for the pure \RbFe{} from Ref.~\cite{smirnov2007}.   Color map at panel (b) shows derivative ${dR}/{dB}$  at corresponding temperatures.  Circles and squares: B-scans (open symbols correspond to the measurements with decreasing field, closed symbols --- to the measurements with increasing field). Triangles: T-scans (open symbols correspond to the data taken on cooling, closed samples --- on heating). Panel (a) includes data from other experiments as follows: green diamond marks zero-field N\'{e}el point determined from magnetization measurements of  Ref.~\cite{smirnov2017}; magenta dotted line shows position of NMR temperature and field scans from Ref.~\cite{sakhratov}, magenta stars along this line (1.47~K and 2.25~K) marks positions of NMR spectra interpreted in terms of the ``fan'' phase with inter-layer disorder, red star (2.8~K) mark position of NMR relaxation rate  $1/T_1$ peak; vertical line at $T=1.3$~K  marks field range studied in low-temperature ESR study of Ref.~\cite{smirnov2017}; vertical lines at 2.8--3.5~K show possible plateau region from high-field magnetization study of Ref.~\cite{smirnov2017}.  }\label{fig:15}
\end{figure}

The observed temperature evolution of the torque-meter response curve $R(B,T)$ for \RbKFe{(1-x)}{x}{} sample with high dilution level $x=15$\% is \emph{qualitatively} different from that of the pure \RbFe{} and the $x=7.5$\% potassium-diluted sample. Examples of the torque-meter response curves are shown in Fig.~\ref{fig:15-data}. The increased degree of structural disorder leads to further smearing of the phase transitions, however certain kink-like features can be distinguished both on the B-scans and on the T-scans of torque-meter response function $R(B,T)$.

Low-temperature B-scans (see upper row in Fig.~\ref{fig:15-data}) demonstrate \emph{ a single kink} at a field of approximately 5---6~T, which is about the field of the ``1/3-plateau'' phase in pure \RbFe{} and in \RbKFe{1-x}{x}{} with low potassium-dilution level. At higher temperatures (around 2.5~K) an extended fragment with a reduced slope appears on $R(B)$ curves, with \emph{two kinks} at its left and right edges (see middle row in Fig.~\ref{fig:15-data}). T-scans (see bottom row in Fig.~\ref{fig:15-data}) also demonstrate a fragment with different slope of the $R(T)$ curve around 2.5---3.0~K.

Positions of these kinks with appropriate error bars accounting for smearing of the transitions are collected on the magnetic phase diagram (Fig.~\ref{fig:15}). This phase diagram is qualitatively different from those in Figs.~\ref{fig:pure},~\ref{fig:7p5}: there is no finite-width plateau-like phase at low temperatures and the plateau-like phase with reduced slope of $R(B)$ curves revives upon heating above 2~K. This plateau-like phase has upper and lower borders in temperature and field dimensions. It seems reasonable to assume, by analogy with the    $x=7.5$\% case, that B-scans overestimate the lower boundary of the plateau-like phase ``cutting'' it tangentially around 2.7~K.

Thus, the phase diagram of the highly diluted \RbKFe{(1-x)}{x}{} sample with $x=15$\% demonstrates suppression of the plateau phase by impurity-induced disorder at low temperatures and revival of the plateau-like phase at higher temperature.

\section{Discussion}

Our experimental results on the magnetic phase diagrams of \RbKFe{1-x}{x}{} can be summarized as follows. Magnetic phase diagram of pure \RbFe{} at $\vect{H}\perp c$ includes fluctuation-stabilized collinear ``1/3-plateau'' phase (``uud'' phase) which has a finite width even at the lowest temperatures (Fig.~\ref{fig:pure}). Potassium dilution introduces static ``frozen'' disorder, which competes with dynamic fluctuations and smears phase transitions \cite{maryasin}. However, characteristic changes of the torque-meter response can be identified and phase boundaries of plateau (or plateau-like) phase can be determined. At the low dilution level  $x=7.5$\%  plateau phase remains robust and is observed down to the  lowest temperatures (Fig.~\ref{fig:7p5}). At the high dilution level $x=15$\% we observe the disappearance of the finite-width plateau phase at low temperatures below 2~K and the revival of the plateau-like phase in a certain field range between 2~K and the N\'{e}el point (Fig.~\ref{fig:15}).

This behavior agrees qualitatively with theoretical predictions \cite{maryasin}. While dynamic fluctuations (both zero-temperature quantum fluctuations and finite temperature thermal fluctuations) favor collinear ordering and thus stabilize collinear ``uud'' phase over the plateau region, the ``frozen'' static disorder acts in the opposite direction. Thus, at a certain disorder level the effect of static disorder can overcome the effect of fluctuations at low temperatures. As temperature increases, thermal fluctuations add weight to the stabilization of collinear phase, which can result in the revival of the ``1/3-plateau'' phase.

Evidences of such an evolution were reported in high-field magnetization study of Ref.~\cite{smirnov2017}: plateau-like phase in $x=15$\% sample was marked by small decrease of $dM/dH$ which was visible from 2.8 to 3.5~K but disappeared at 1.8~K.  Our measurements yield broader area on the phase diagram corresponding to the plateau-like phase (see Fig.~\ref{fig:15}), the difference can be naturally explained to the high noise and loss in the sensitivity of pulsed high-field experiments of Ref.~\cite{smirnov2017} below  4~T.

Low-temperature ESR data for $x=15$\% sample \cite{smirnov2017} do not show any trace of spin excitations modes softening similar to that observed at the plateau region in pure \RbFe{} at $T=1.3$~K. This is in agreement with our results pinning down the low-temperature border of the plateau-like phase at approximately 2~K.

An attempt to decipher the structure of the ordered  phase in \RbKFe{1-x}{x}{} was performed in NMR experiment of Ref.~\cite{sakhratov}. Transformation of the NMR spectra on crossing from paramagnetic to magnetically ordered phase was detected and the ordered phase was identified via NMR line-shape modeling as a ``fan'' phase with partial inter-layer disorder. However, one can note that the NMR spectra used in Ref.~\cite{sakhratov} for line-shape modeling (their position in the $(BT)$ plane is marked by magenta stars in Fig.~\ref{fig:15}) are outside of the plateau-like phase determined in the present work. Suggested plateau-like phase boundaries correspond to broadening and continuous transformation of NMR spectra with peak of $T_1^{-1}$ being within the plateau-like phase (see red star in Fig.~\ref{fig:15}).

Summing up, we conclude that the determined phase boundaries of the plateau-like phase in highly diluted \RbKFe{1-x}{x}{} are in reasonable agreement with known results from other measurement techniques. Our study allowed to trace out these phase boundaries systematically.

The microscopic structure of the plateau-like phase can not be established from the present experiments. Magnetization measurements of Ref.~\cite{smirnov2017} demonstrate that $dM/dH$ decreases by less than 10\% within the plateau-like phase. NMR spectra \cite{sakhratov} within the plateau-like phase do not show characteristic multi-peak structure typical for the ``true'' collinear ``uud'' phase observed in pure \RbFe{}. Thus, one could expect superposition of ``uud'' ordering pattern with other order parameters within the plateau-like phase, possibly additionally complicated by the inter-layer disorder as is observed within the ``fan'' phase \cite{sakhratov}.

Now we briefly address tentative phase transition line at approximately 5~T below 2~K (see Fig.~\ref{fig:15}). This line is marked by quite a sharp kink on the $R(B)$ curve. Interestingly, $R(B)$ curves taken at different temperatures approach this line from lower fields with approximately the same slope $dR/dB$ as shown by the color map in Fig.~\ref{fig:15}. Torque-meter data alone can not tell much about this feature. Moreover, we can not reliably distinguish a smeared phase transition from a crossover. Simple three-sublattice modeling (Fig.~\ref{fig:model}) yields similarly positioned ``inverted-T'' line at a crossover from ``anti-Y'' to ``fan'' ordering pattern.

It is an open question both for the theory and for other experimental techniques whether this crossover rise to a phase transition in 3D or not. Low-temperature ESR experiment of Ref.~\cite{smirnov2017} did not report any features which could be related to the possible phase transition at approximately~5~T. Note also, that $M=M_\textrm{sat}/3$ along the  ``inverted-T'' line --- the same value as over the magnetization plateau. Thus, the splitting of this line into the extended range of fields over the magnetization plateau looks as a reasonable evolution route.

Considering diluted system one has to keep in mind possible formation of finite-size spin clusters. Order-by-disorder mechanism affects magnetization process even for triangular cluster  \cite{glazkov-cluster} yielding  drop of differential susceptibility  $dM/dB$ at approximately 1/3 of saturation field. Such a drop of   $dM/dB$ resembles behavior of the torque-meter response curve $R(B,T)$ within the plateau-like phase of $x=15$\% potassium-diluted sample to some extent. However, there are significant qualitative differences: (i) finite-size cluster model \cite{glazkov-cluster} does not predict a single-kink $M(B)$ behavior observed at low temperatures; (ii) two-kinks $R(B,T)$ behavior features well-defined upper and lower boundaries, while drop of  $dM/dB$ in the finite-size cluster model is smooth. Thus, we can conclude, that the phase boundaries shown in Fig.~\ref{fig:15} correspond to the real phase transitions in macroscopic magnet, but not to the effects of the formation of finite-size spin clusters at high dilution level.

\section{Conclusions}

We present results of the detailed magnetic phase diagram study of diamagnetically diluted \RbKFe{1-x}{x}{}. We observed the effect of revival of the collinear phase under competing dynamic and static disorder, as  predicted theoretically \cite{maryasin}.

\acknowledgements

The work was supported by Russian Science Foundation Grant 22-12-00259-$\Pi$. The authors thank Prof.~A.~Smirnov, Prof.~L.~Svistov and Dr.~S.~Sosin (Kapitza Institute) for stimulating discussions.

\end{document}